# Enhanced absorption per unit mass for infrared arrays using subwavelength metal-dielectric structures

AVIJIT DAS[1] AND JOSEPH J. TALGHADER[1,*]

[1]*Department of Electrical and Computer Engineering, University of Minnesota, Minneapolis, MN 55455, USA.*
*joey@umn.edu*

**Abstract:** The absorption-to-mass ratio of the infrared arrays is enhanced to ~1.33−7.33 times larger than the previously reported structures by incorporating two design characteristics: first, the coupling of evanescent fields in the air gaps around pixels to create effectively larger pixel sizes, and, second, the use of guided mode resonance (GMR) within the subwavelength metal-dielectric gratings. The bilayer Ti-$Si_3N_4$ gratings achieve broadband long-wave infrared (LWIR, λ ~ 8−12 μm) absorption by the combined effects of free carrier absorption by the thin Ti films and vibrational phonon absorption by the thick $Si_3N_4$ films. In the presence of GMR, this broadband absorption can be enormously enhanced even with low fill factor subwavelength grating cells. Further, the spacing and design of the cells can be modified to form a pixel array structure that couples the light falling in the air gaps via evanescent field coupling. Calculations are performed using the finite difference time domain (FDTD) technique. Excellent broadband absorption is observed for the optimized arrays, yielding maximum absorption of 90% across the LWIR and an average absorption-per-unit-mass (absorption/mass) per pixel of $3.45 \times 10^{13}$ $kg^{-1}$.

## 1. Introduction

Achieving near-perfect LWIR absorption with wide bandwidth has become a critical goal in the design and implementation of infrared thermal and imaging devices [1−4]. With the advent of science and technology, it has become possible to design LWIR broadband absorbers with ultra-thin sub-wavelength thicknesses [5−8]. Furthermore, nearly perfect absorbers have also been studied for their use in bio-sensing [9, 10], photodetection [11], and thermal emission based-cooling [12−15].

In recent years, relevant research has focused on the design and development of perfect LWIR absorbers with small amounts of material [16−18]. Maximizing absorption in a system with low thermal mass is crucial, for example, in microbolometers and other thermal detectors. In a traditional microbolometer, one attempts to reduce mass in order to achieve a lower thermal time constant (in other words, faster response) [3, 17, 19]. For high performance radiation-limited thermal detectors, the concept is a bit more subtle. The minimum mass of a detector is limited by the magnitude of intrinsic thermal fluctuations that can be tolerated on the device; therefore, any mass below this is counterproductive. With this given thermal mass, one then wishes to maximize the area of the detector so that the thermal conductance, which is dependent on the area for radiation-limited devices, can be maximized, reducing the time constant.

In order to achieve high LWIR absorption with low mass, a number of structures can be investigated, such as metamaterial absorbers [20−23] and plasmonic composites [24, 25]. Typically metamaterials are comprised of three distinct layers, (a) a bottom continuous metal layer, (b) an insulator spacing layer, and (c) a top subwavelength metal patch layer [5, 6]. In such a metal-insulator-metal (M-I-M) configuration, maximum light coupling is realized by perfect impedance matching with the incident medium, resulting in minimum (near zero) reflection. However, perfect impedance matching can only be obtained over a narrow frequency range, specifically with uniformly sized metal patches supporting a single resonant band [26, 27]. To broaden the bandwidth, one approach is to tailor the size of the planar metal patches in a single unit cell (i.e., multiple planar resonators in a unit cell), resulting in a superposition of multiple resonance bands [6]. In addition to increasing the absorption area, this approach has a couple of drawbacks: (1) larger volume of material in a period, resulting in smaller absorption per unit mass, and (2) the magnetic response may not be simultaneously tuned for all resonator patches in a single unit, resulting in an imperfect impedance matching and hence, imperfect optical absorption [26, 27]. Another convenient approach is to pattern the top layer with alternating metal-dielectric (i.e. metal-insulator, M-I) thin films vertically stacked in a graded fashion, resulting in anisotropic sawtooths [28] or trapezoidal pyramids [8, 29]. The vertically stacked films allow multiple resonant frequencies to be spaced adjacent to one another with smaller intervals among the resonance bands than the planar resonators, resulting in a reduction of the quality factor over a wide bandwidth. However, maximum absorption with low quality factor can only be achieved when a large number of metal-dielectric films can be stacked in a period, eventually increasing the total volume and decreasing absorption per unit mass. For example, a graded trapezoidal pyramid requires a thickness of 6 μm to maintain perfect broadband absorption in the range of wavelengths from 10 μm to 30 μm [29]. Other metamaterial approaches for perfect broadband LWIR absorption include hyperbolic metamaterials (doped-undoped semiconductor stacks) [30, 31] and unpatterned metal-dielectric pairs [32], which require even larger amount of materials per pixel or period than M-I-M structures. In case of plasmonic structures, perfect LWIR absorption is typically observed over a narrow bandwidth, which can be attributed to their highly dispersive resonance characteristics [5, 9, 16]. A few works have reported broadband LWIR absorption through the superposition of multiple resonance bands at the cost of larger volumes of plasmonic structures [24, 25], yielding again larger amount of material. There is hardly, if any, work reported for broadband absorption over the LWIR transparency window that realizes near-perfect absorption with minimum material.

In this paper, we present a broadband LWIR pixel array structure that maximizes absorption per unit mass by utilizing two-dimensional subwavelength gratings with guided mode resonance (GMR) in the pixel structure, and further designing the array so that gaps between pixels act as part of the absorption structure via evanescent field coupling. In recent years, GMRs have been exploited to design LWIR filters [33, 34] by virtue of their simple structures and nearly 100% diffraction efficiency. In our structure, we use lossy metal-dielectric (i.e., Ti-$Si_3N_4$) bilayer waveguide gratings with low fill factors (large hole fractions) that enable evanescent field coupling inside the open holes. Instead of the conventional M-I-M configuration used in many metamaterials, we design with only a metal-dielectric (two layer) grating, resulting in enhanced absoption per unit mass. We design and optimize a GMR pixel array (i.e., pixels surrounded by periodic air gaps), where each pixel is formed by the low fill factor Ti-$Si_3N_4$ grating (or grid unit) cells. We calculate the broadband LWIR absorption of the GMR pixels using the finite difference time domain technique (FDTD). Finally, we calculate and analyze the absorption per unit mass of the resulting pixel.

## 2. Theoretical design of the GMR pixel with subwavelength gratings

*2.1 Metal-dielectric subwavelength grating cells*

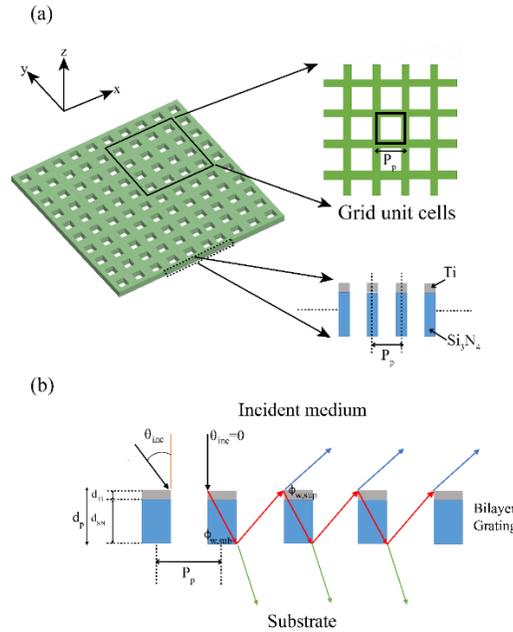

**Fig. 1.** (a) Schematic illustration of grid unit cells constructed from the metal-dielectric (Ti-Si$_3$N$_4$) waveguide gratings. Period of a grid unit cell (P$_P$) is highlighted, along with Ti and Si$_3$N$_4$ layers. (b) Incidence of light in the Ti-Si$_3$N$_4$ waveguide gratings; some portion will be reflected (blue arrows) and transmitted (green arrows), and the rest will propagate laterally inside the grating structure, resulting in trapping and absorption of light (red arrows). The thicknesses of the Ti and Si$_3$N$_4$ layers in the grating are presented as d$_{Ti}$ and d$_{SN}$, respectively. The total thickness is shown as d$_p$, i.e., d$_p$= d$_{Ti}$+ d$_{SN}$. $\phi_{w,sup}$ and $\phi_{w,sub}$ present phase shifts due to the total internal reflection at the waveguide-incident medium and waveguide-substrate interfaces.

Figure 1(a) shows the metal-dielectric (Ti-Si$_3$N$_4$) waveguide gratings in the form of grid unit cells. The period of the grid unit cells (P$_P$) is explicitly shown along with the metal and dielectric layers. Such optical gratings can support resonant waveguide modes depending on a number of parameters, including the complex dielectric permittivity, the thickness of the grating layers, and the hole fraction of the grating structure [35]. To ensure enhanced LWIR broadband absorption with small mass, lossy mateials need to be chosen so that both refraction and extinction can contribute to the guided mode excitation. The mechanism of enhanced absorption due to lossy dielectrics has already been reported in previous articles [4, 36, 37]. Compatibility with common microfabrication technology makes SiO$_2$ and Si$_3$N$_4$ two of the simplest choices for lossy dielectrics in the LWIR. SiO$_2$ has a very sharp absorption peak at ~9.22 μm, resulting in highly dispersive characteristics in the LWIR [38]. Si$_3$N$_4$ has multiple vibrational phonon peaks in between 8–12 μm, which correspond to a comparatively moderate dispersion and broader absorption profile in the LWIR [39]. For this reason, we choose Si$_3$N$_4$ as the lossy dielectric. The addition of a thin metal layer (absorbing in the LWIR) on top of the dielectric layer can significantly reduce the quality factor and broaden the entire absorption band [40]. A number of metal films e.g., Ti , Cr and Ni have already been used for this purpose due to their similar optical properties [41 – 43]. However, Cr and Ni have densities approximately two times larger than Ti [44], which would increase the overall amount of mass. Therefore, we choose Ti as the lossy thin metal film to ensure enhanced absorption with smaller amount of mass. At this point, we should note that for some applications, such as thermal detectors, the optimal performance metric is absorbance per unit thermal mass per unit volume instead of absorbance per unit mass. In these cases, the product of the density and specific heat for Ti, Cr, and Ni would be technically more accurate than density alone. In any case, Ti is also the best choice by this metric: the density-specific heat products of Cr and Ni are 40.3% and 65.8% higher than that of Ti, respectively.

Figure 1(b) shows 1D illustration of the bilayer waveguide gratings with incidence of light ($\theta_{inc}$). Some portion of the light will reflect (blue arrows) or transmit (green arrows), and the rest will propagate inside the waveguide grating (red arrows), resulting in total internal reflection at the waveguide-incident medium and waveguide-substrate interfaces. At the guided mode resonant condition, the penetration of the light in the substrate or incident medium can be greatly reduced by forcing the diffracted light to propagate laterally and thus increasing the optical path inside the waveguide grating, resulting in maximum absorption. To ensure guided mode excitation, the grating thickness needs to be optimized with different hole fractions (i.e. duty cycles or fill factors). This requires an effective medium approximation in the waveguide grid unit cells. It should be noted that the dimensions of our subwavelength grating cells are well below the wavelength of the light in consideration, therefore effective medium approximation can be applied for the subwavelength domain. Fig. 2(a) presents the schematic of a grid unit period with the complex dielectric permittivities:

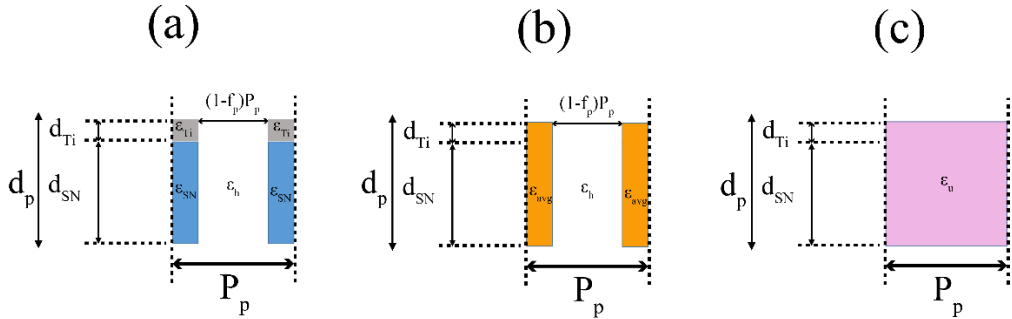

**Fig. 2.** Equivalent permittivity approximation of a grid unit cell. (a) Schematic illustration of a grid unit period with the complex dielectric parameters; $\varepsilon_{Ti}$, $\varepsilon_{SN}$, and $\varepsilon_h$ for Ti, $Si_3N_4$, and the hole (open) area, respectively. $f_p$ and $P_p$ correspond to the linear duty cycle and period of the unit cell, respectively. (b) As in part (a) but with an average permittivity $\varepsilon_{avg}$ of the Ti- $Si_3N_4$ bilayer. (c) As in parts (a) and (b) but with an equivalent homogeneous permittivity $\varepsilon_u$ of the grid unit period.

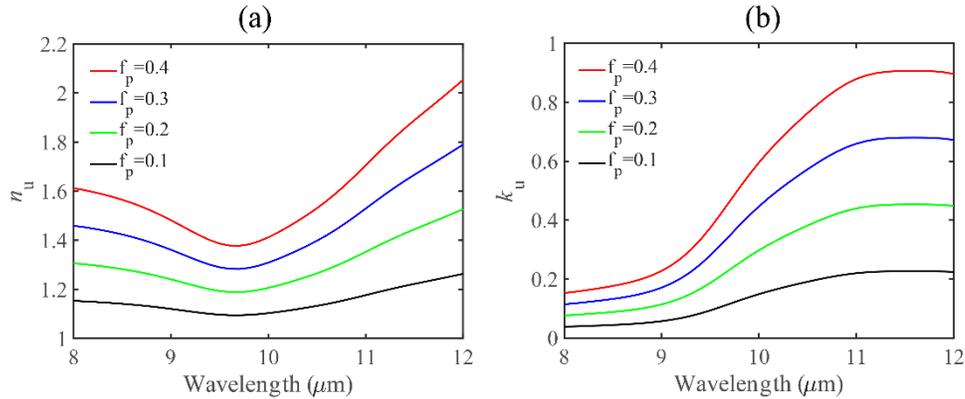

**Fig. 3**. Refractive index ($n_u$) and extinction coefficient ($k_u$) of the equivalent homogeneous grid unit cell. The thicknesses are assumed as $d_{Ti}=0.2d_p$ and $d_{SN}=0.8d_p$. $f_p$ corresponds to the linear duty cycle of the unit cell, which in turn defines the fill factor (fractional area filled by solid material) and hole fraction.

$\varepsilon_{Ti}$, $\varepsilon_{SN}$, and $\varepsilon_h$ for Ti, Si$_3$N$_4$, and hole (open) area, respectively. $f_p$ corresponds to the linear duty cycle of the unit cell, which in turn defines the fill factor (fractional area filled by solid material) and hole fraction. The thicknesses of Ti and Si$_3$N$_4$ layers in the unit cell are presented as $d_{Ti}$ and $d_{SN}$, respectively. The total thickness is shown as $d_p$, i.e., $d_p = d_{Ti} + d_{SN}$. Let us assume $d_{Ti}=0.2d_p$ and $d_{SN}=0.8d_p$, considering maximum light coupling inside the dielectric layer (i.e., absorption dominated by dielectric loss) to avoid any unnecessary thermal heating due to metals [45]. The metal-dielectric bilayer grating can be approximated as a single layer with an average permittivity calculated as [46–50],

$$\varepsilon_{avg} = \left( \frac{d_{Ti}}{d_p} \frac{1}{\varepsilon_{Ti}} + \frac{d_{SN}}{d_p} \frac{1}{\varepsilon_{SN}} \right)^{-1} \tag{1}$$

Fig. 2(b) shows the metal-dielectric bilayers approximated by a single effective layer with permittivity $\varepsilon_{avg}$. After single layer approximation, the hole ratio can be incorporated in the grid unit period to form an equivalent medium by the effective medium approximation as [47, 51–55],

$$\varepsilon_u = f_p \varepsilon_{avg} + (1-f_p)\varepsilon_h = f_p \left( \frac{d_{Ti}}{d_p} \frac{1}{\varepsilon_{Ti}} + \frac{d_{SN}}{d_p} \frac{1}{\varepsilon_{SN}} \right)^{-1} + (1-f_p)\varepsilon_h \tag{2}$$

where $\varepsilon_u$ is the equivalent homogeneous permittivity of the grid unit cell (in Fig. 2(c)). Note that for air, $\varepsilon_h = 1$. Fig. 3 shows the refractive index $n_u$ and extinction coefficient $k_u$ calculated from $\varepsilon_u$ (i.e., $n_u + ik_u = \sqrt{\varepsilon_u}$) with different values of $f_p$. Please note that we focus on investigating enhanced absorption with a minimum amount of material, therefore lower values of $f_p$ have been considered in our analysis. From Fig. 3, it is quite evident that a lower $f_p$ would decrease both $n_u$ and $k_u$ values due to a larger hole ratio. Since the silicon nitride layers are much thicker than those of Ti, $n_u$ and $k_u$ profiles of the homogeneous grid cell are much dominated by those of silicon nitride [56] (please see Supplement 1). As a result, dielectric losses become prominent over much of the spectral range. Please note that the effective medium approximation has been previously reported in analyzing the optical absorption of the grating and periodic array structures [4, 57–59]. The effective medium cells show optical absorption nearly close to those of the actual structures (for example, an average relative error of ~3% found while analyzing in between 0.35−0.9 μm [57] and ~8.5% found while analyzing in between 7−14 μm [4]).

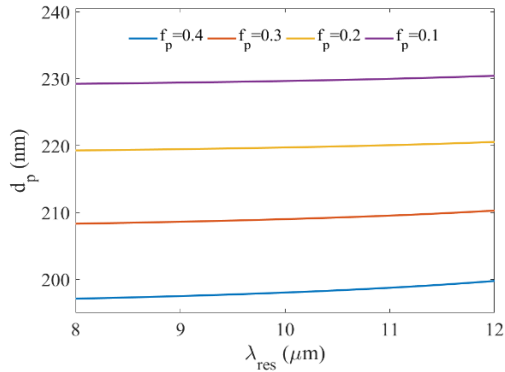

Fig. 4. Based on guided mode resonance theory, the thickness of the grid unit cell is optimized for diffraction efficiency with different duty cycles $f_p$. The *x*-axis shows the guided resonant wavelengths in the LWIR. Normal incidence of light ($\theta_{inc}=0°$) and $P_p=1$ µm are considered with first order diffraction (m=±1).

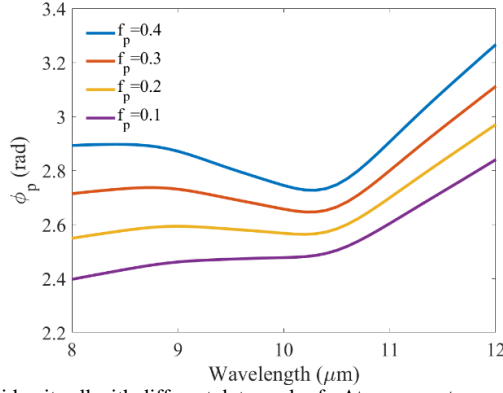

**Fig. 5**. Phase shifts of the grid unit cell with different duty cycles $f_p$. At a resonant wavelength (e.g., $\lambda_{res}$= 10.5 μm), the optimum thickness $d_p$ for different duty cycle $f_p$ are taken from Fig. 4. Normal incidence of light ($\theta_{inc}$=0°) and $P_p$=1 μm are considered with first order diffraction (m=±1).

To optimize the diffraction efficiency based on GMR, the propagation waveguide modes in the grid unit cells with homogeneous permittivity $\varepsilon_u$ can be coupled with the diffraction grating equation as [35, 60],

$$\tan(\kappa_i d_p) = \frac{\varepsilon_u \kappa_i (\varepsilon_{sub} \gamma_i + \varepsilon_{sup} \delta_i)}{\varepsilon_{sub} \varepsilon_{sup} \kappa_i^2 - \varepsilon_u^2 \gamma_i \delta_i} \quad (3)$$

where

$$\begin{cases} \kappa_i = \sqrt{\varepsilon_u k_0^2 - \beta_i^2} \\ \gamma_i = \sqrt{\beta_i^2 - \varepsilon_{sup} k_0^2} \\ \delta_i = \sqrt{\beta_i^2 - \varepsilon_{sub} k_0^2} \\ \beta_i = k_0 \left( \sqrt{\varepsilon_{sup}} \sin \theta_{inc} + m \frac{\lambda}{P_p} \right) \\ k_0 = \frac{2\pi}{\lambda} \end{cases} \quad (4)$$

where λ is the wavelength, $\theta_{inc}$ is the angle of incidence (at normal incidence, $\theta_{inc}$=0°), m is the diffraction order, $\varepsilon_{sup}$ and $\varepsilon_{sub}$ are the permittivity of the incident medium and substrate, respectively (for air, $\varepsilon_{sup} = \varepsilon_{sub} = 1$). Using this model, it is possible to evaluate the grid unit cell thickness $d_p$ optimized for diffraction efficiency (based on GMR theory) across the LWIR (i.e., 8–12 μm) with different duty cycles $f_p$. Fig. 4 presents the optimum thicknesses $d_p$ of the grid unit gratings for different duty cycles $f_p$, with the assumption of $\theta_{inc}$=0° (normal incidence), m=±1 (first-order diffraction), and $P_p$=1 μm. The x-axis corresponds to the guided resonant wavelengths in the LWIR. Please note that only the transverse magnetic (TM) polarization is calculated since 2D symmetric groove or pillar gratings (e.g., shown in Fig. 1(a)) at normal incidence are polarization independent [61−65], and can be analyzed with only TM (or only TE) polarized light [62, 63]. From Fig. 4, it is evident that for each duty cycle, the optimum thickness values are almost same over the whole LWIR range, which can ensure resonant absorption with a wide bandwidth. In addition, a lower duty cycle corresponds to a larger grating thickness for guided mode excitiation, which has been oberserved in previously reported waveguide gratings [66].

At a particular optimum thickness, we can calculate the phase shifts of the diffracted light inside the grid unit cells to study the resonance quality (high-Q or low-Q) of the guided modes. As seen in Fig. 1(b), light traveling inside the waveguide grating encounters phase shifts that are characteristic of the various propagation modes, as well as, total internal reflections at the substrate-waveguide and incident medium-waveguide interfaces. Therefore, the total phase shift of the diffracted light inside the grid cell can be approximated as [67, 68],

$$\phi_p = 2k_0 \left| \sqrt{\varepsilon_u - \varepsilon_{eff}} \right| d_p + \phi_{w,sup} + \phi_{w,sub} \tag{5}$$

where

$$\begin{cases} \sqrt{\varepsilon_{eff}} = \dfrac{\beta_i}{k_0} = \sqrt{\varepsilon_{sup}} \sin\theta_{inc} + m\dfrac{\lambda}{P_p} \\ \phi_{w,sup} = -2\tan^{-1} \left| \dfrac{\varepsilon_u}{\varepsilon_{sup}} \dfrac{\sqrt{\varepsilon_{sup} - \varepsilon_{eff}}}{\varepsilon_u - \varepsilon_{eff}} \right| \\ \phi_{w,sub} = -2\tan^{-1} \left| \dfrac{\varepsilon_u}{\varepsilon_{sub}} \dfrac{\sqrt{\varepsilon_{sub} - \varepsilon_{eff}}}{\varepsilon_u - \varepsilon_{eff}} \right| \end{cases} \tag{6}$$

where $\varepsilon_{eff}$ is the effective permittivity of the guided mode propagating through the waveguide gratings (obtained from Eq. (4)). In order to demonstrate the first order diffracted phase shifts in between 8−12 μm, the permittivity $\varepsilon_u$ for different duty cycles $f_p$ can be approximated from Eq. (2). In addition, a central resonant wavelength (e.g., 10.5 μm) can be considered for choosing the optimum grating thickness $d_p$ with different values of $f_p$. From Fig. 4, we took $d_p$=230, 220, 210 and 198 nm for $f_p$=0.1, 0.2, 0.3 and 0.4, respectively. Fig. 5 shows the phase shifts of the grid unit calculated at different $f_p$ values. Apart from a slight notch at the resonant wavelength, the phase curve shows a very small and gradual change (~4.7% of 2π) throughout the LWIR (near fundamental mode resonance). It should be noted that the amount of phase shift is very sharp and abrupt if the resonance band is small (i.e., high-Q resonance) [68, 69], whereas the phase shift is slow and gradual if the resonance band is large (i.e., low-Q resonance) [70]. Therefore, the phase curves in Fig. 5 clearly suggest a very low-Q resonance in the optimized grid units.

*2.2 Metal-dielectric grating pixel*

Using the optimized Ti-Si$_3$N$_4$ grid unit cells, we can design the grating pixel array with enhanced absorption per unit mass. Fig. 6 shows the metal-dielectric grating pixel array (period $P_t$) formed by the grid unit cells (period $P_p$) and air gaps ($g_t$) between adjacent pixels.

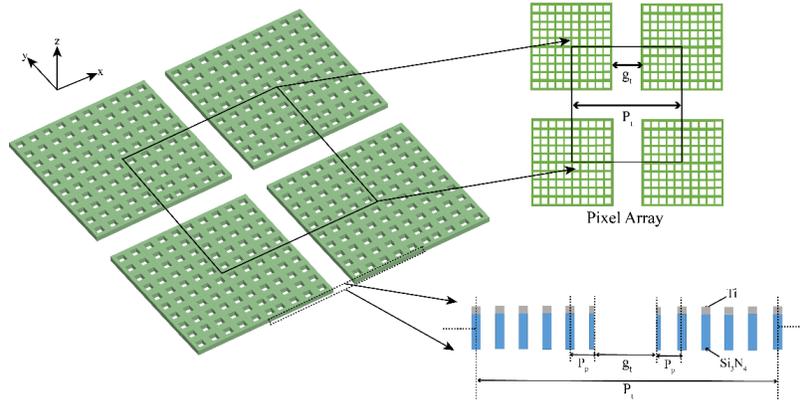

**Fig. 6.** Schematic illustration of pixel array structure constructed from the grid unit cells. Period of a grating pixel ($P_t$) is highlighted, along with the grid unit cells and air gap ($g_t$) between adjacent pixels.

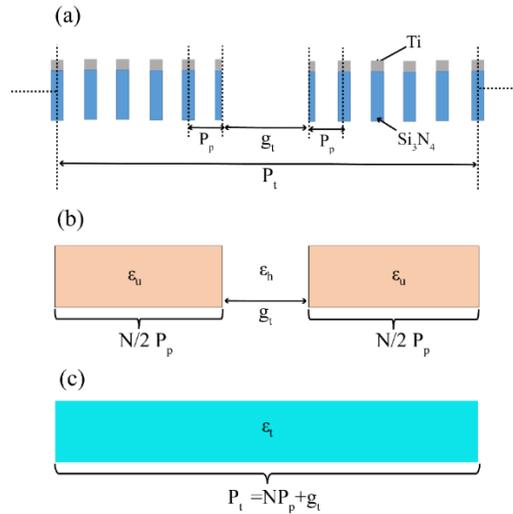

**Fig. 7.** Analysis of effective permittivity of a periodic pixel with subwavelength gratings. (a) Schematic illustration of a pixel period ($P_t$) with grid unit cells (period $P_p$) and air gap ($g_t$) between adjacent pixels. (b) Equivalent permittivity $\varepsilon_u$ of the grid unit cells is shown along with the air gap permittvity $\varepsilon_h$. A total of N grid unit cells are present in the pixel period. (c) Pixel period with an effective permittivity $\varepsilon_t$

Guided mode excitation in the pixel structure can be properly optimized when analyzed with an effective optical permittivity. Fig. 7 presents the effective permittivity analysis of the 1D grating pixel. In Fig. 7(a), N bilayer grid unit cells are symmetically arranged on both sides of the air gap $g_t$ to form the pixel period. In Fig. 7(b), the bilayer unit cells are replaced by the equivalent homogeneous cells with permittivity $\varepsilon_u$ (from Fig. 2). Therefore, the grating pixel takes the form of a periodic structure, with homogenous medium $\varepsilon_u$ (filling portion) and air gap medium $\varepsilon_h$ (hole portion). Based on this form, the pixel period $P_t$ can be defined as $P_t = NP_p + g_t$. In Fig. 7(c), the pixel period is approximated by an effective permittvity $\varepsilon_t$, which can be calculated as [47, 71, 72],

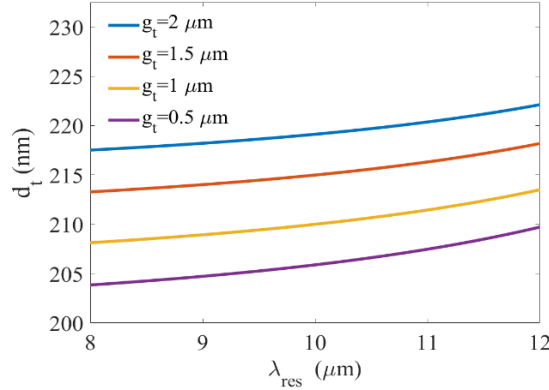

**Fig. 8** Based on guided mode resonance theory, the thickness of the grating pixel is optimized for diffraction efficiency with different air gap distances $g_t$. The x-axis shows guided mode resonant wavelengths in the LWIR. Normal incidence of light ($\theta_{inc}=0°$), $P_p=1$ µm, $f_p=0.2$, and N=10 are considered with first order diffraction (m=±1).

$$\varepsilon_t = \varepsilon_h + \Delta\varepsilon f_t$$
$$\begin{cases} \Delta\varepsilon = \varepsilon_u - \varepsilon_h \\ f_t = \dfrac{NP_p}{NP_p + g_t} \end{cases} \quad (7)$$

where $f_t$ is the linear duty cycle of the pixel period. It should be noted that Eq. (7) is validated by the effective grating theory [71 − 73], which applies for gratings with subwavelength thickness and period less than or comparable to the illumination wavelength (please see Supplement 1).

In order to obtain enhanced absorption per unit mass, the grating pixel needs to be designed with low fill factor grid units (i.e., small amounts of material) forming a large period (i.e., large absorption area). Therefore, in Eq. (7), the permittivity $\varepsilon_u$ of the grid unit cells must be calculated using a low duty cycle $f_p$. Considering good optical properties from Fig. 3 (e.g. effective refractive index close to unity, resulting in minimum reflection) and a requirement of minimum material, we use grid unit cells with $f_p=0.2$ and $P_p=1$ µm in our analysis. Please note that although $f_p=0.1$ (and other lower values) would require less material, free-standing periodic grid units with $f_p=0.1$ and $P_p=1$ µm have significant fabrication limitations [17]. To form a large pixel, we consider 10 grid units inside the pixel period $P_t$ (i.e., N=10). By increasing the periodic air gap $g_t$, the pixel period can be further increased, which, in turn, requires proper optimization of the pixel thickness based on GMR theory. To ensure maximum evanescent field coupling inside the air gaps, pixel thickness $d_t$ can be optimized for diffraction efficiency across the LWIR (i.e., 8−12 µm) when the grid unit permittivity $\varepsilon_u$ and period $P_p$ are replaced with pixel permittivity $\varepsilon_t$ and period $P_t$ in Eqns. (3) and (4), respectively. Fig. 8 shows the optimum pixel thicknesses $d_t$ for different air gaps $g_t$, with the assumptions of $\theta_{inc}=0°$ (normal incidence), m=±1 (first-order diffraction), $P_p=1$ µm, $f_p=0.2$, and N=10. The x-axis indicates guided mode resonant wavelengths in the LWIR. It can be clearly observed that an increase in gap $g_t$ (i.e., decrease in $f_t$) would require a larger pixel thickness $d_t$ to compensate for the overall amount of material required for the resonant coupling. This is similar to the case seen in Fig. 4. Moreover, for a particular $g_t$, the optimum thicknesses are close to one another over the LWIR range (with a deviation of only ~3%).

Similarly to the grid units, the resonance quality (high-Q or low-Q) of the guided modes inside the pixel can be studied by calculating the phase shifts of the light diffraction. By replacing permittivity $\varepsilon_u$ with $\varepsilon_t$, and period $P_p$ with $P_t$ in Eqns (5) and (6), the phase information of the pixel structure can be extracted. Note that at the resonant wavelength (e.g., 10.5 µm), the optimum pixel thicknesses $d_t$ for different air gaps $g_t$ are used to calculate the phase shifts. The values of $d_t$ are taken from Fig. 8. The other specifications are kept fixed ($\theta_{inc}=0°$, $P_p=1$ µm, $f_p=0.2$, and N=10), and the phase calculations of the pixel structure for different air gaps are presented in Fig. 9. The relatively smooth and small change (~6% of $2\pi$) in slope across the LWIR resonance implies a very low-Q system [70].

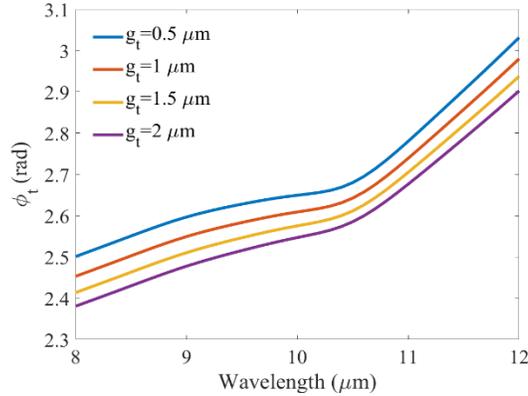

**Fig. 9.** Phase shifts of the grating pixel with different air gaps ($g_t$). At a given resonant wavelength (e.g., $\lambda_{res}$= 10.5 µm), the optimum thickness $d_t$ corresponding to different $g_t$ are taken from Fig. 8 for calculating the phase shifts. The simulated conditions are: normal incidence ($\theta_{inc}$=0°), $P_p$=1 µm, $f_p$=0.2, and N=10 with first order diffraction (m=±1).

## 3. Analysis of optical absorption

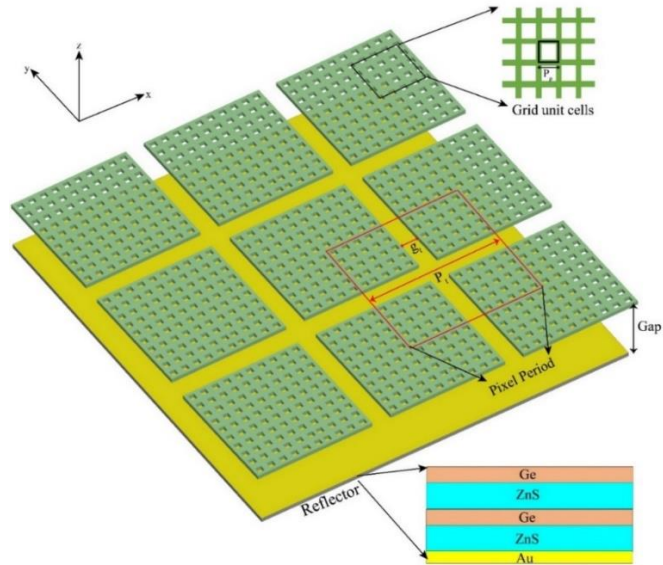

**Fig. 10** Schematic representation of the Ti-Si$_3$N$_4$ grating pixel array with a bottom reflector placed at a gap distance of 2.5 µm. The reflector is made of two pairs of Ge-ZnS layers and a thin 100 nm Au film. The layers are arranged to approximate Bragg mirrors [74], and the thicknesses of the Ge and ZnS layers are taken to be 0.625 µm and 1.136 µm, respectively. With the Au film underneath, the reflector ensures nearly 100% reflection of the LWIR (please see Supplement 1), and therefore, maximum light coupling within the grating pixel.

Figure 10 presents a schematic illustration of our Ti-Si$_3$N$_4$ grating pixel used as a broadband absorber. The pixels are arranged in a periodic fashion along the *x* and *y* directions. A bottom reflector is placed underneath the pixel structure at a gap distance of 2.5 µm, which is a quarter of the wavelength of 10 µm [17]. The reflector is made of two pairs of Ge-ZnS layers and a thin 100 nm Au film. According to the Bragg mirror approximation [74], the thickness of Ge and ZnS layers are taken as 0.625 µm and 1.136 µm, respectively. With the Au film underneath, the reflector ensures nearly 100% reflection of the LWIR (please see Supplement 1), and therefore, maximum light coupling within the grating pixel. Such backside reflectors have already been demonstrated for increasing the GMR absorption in previous studies [75, 76].

To calculate the optical absorption, we solve Maxwell's electromagnetic equations using a finite difference time domain (FDTD) technique in Lumerical Solutions [77]. Assuming a full 3D model, we analyze a pixel period and/or grid unit period with periodic boundary conditions along the horizontal directions (i.e., *x*- and *y*- axis) and with perfectly matched layer (PML) conditions (i.e., no reflection of light from the boundaries) along the vertical direction (i.e., *z*-axis). We assume the incident light is a plane wave directed normally to the surface (i.e., $\theta_{inc}=0°$) and with wavelengths ranging from 8 µm to 12 µm. We take the frequency-dependent optical properties (n and k) of Ti, Si$_3$N$_4$, Ge, ZnS, and Au from previous theoretical and experimental studies [56, 78−81]. In the simulations, we employ a mesh size of 20 nm along the *x*- and *y*-axis (grating plane), and a mesh size of 5 nm along the *z*- axis (thickness). The absorption in the simulations is defined as $A(\lambda) = 1 - R(\lambda) - T(\lambda)$, where R($\lambda$) and T($\lambda$) present the reflection and transmission, respectively.

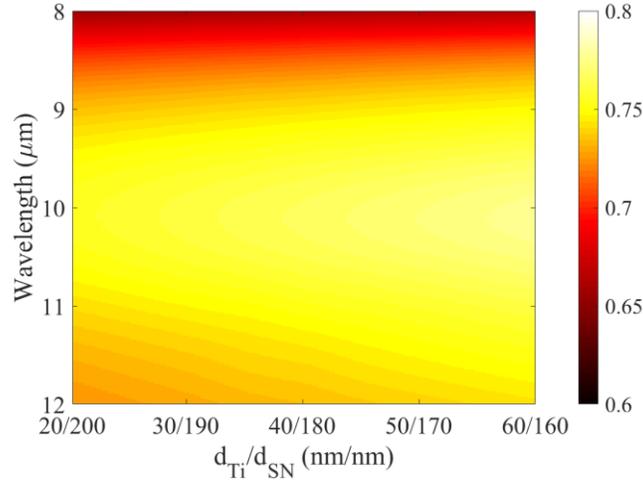

**Fig. 11** Absorption profile of periodic grid unit cells with different thickness combinations of Ti (d$_{Ti}$) and Si$_3$N$_4$ (d$_{SN}$). The duty cycle f$_p$ and period P$_P$ are taken as 0.2 µm and 1 µm, respectively. The total thickness is kept as 220 nm to maintain the GMR condition (from Eq. (3)).

Initially we analyze an optimized grid unit cell with duty cycle f$_p$=0.2 and period P$_p$=1 µm. This would correspond to a fill area factor of 36% (hole area of 64%). From our prior theoretical analysis (Fig. 4), we take the grid unit thickness as 220 nm. While keeping the resonance condition approximately constant with a total thickness of 220 nm, we vary the thicknesses of Ti and Si$_3$N$_4$ layers to study the effect on the optical absorption. Fig. 11 shows the optical absorption for the grid unit period with different thickness combinations of Ti and Si$_3$N$_4$ layers (i.e., d$_{Ti}$/d$_{SN}$). It can be observed that larger Ti thickness gradually broadens the optical absorption, while keeping the resonance in between 9−11 µm. This can be attributed to the large extinction coefficient of Ti causing a decrease in the quality factor since $Q \propto \dfrac{1}{k}$ [40]. In addition, larger Ti layer does not enormously increase the absorption profile (only ~6% increase in maximum absorption), which is indicative of a primarily dielectric-loss-based absorption. Such absorption characteristics (dominated by dielectric loss) in 2D symmetric gratings (formed by metal and dielectric) have been previously reported [4, 36], in which metals layers are found to contribute ~5−8% to the total absorption in the LWIR. We choose 50 nm Ti and 170 nm $Si_3N_4$ for our grating pixel to obtain a uniformly high absorption profile across the LWIR. It should be noted that these thicknesses correspond to $d_{Ti}=0.22d_p$ and $d_{Ti}=0.78d_p$, which are almost same as our theoretical analysis (i.e., $d_{Ti}=0.2d_p$ and $d_{SN}=0.8d_p$).

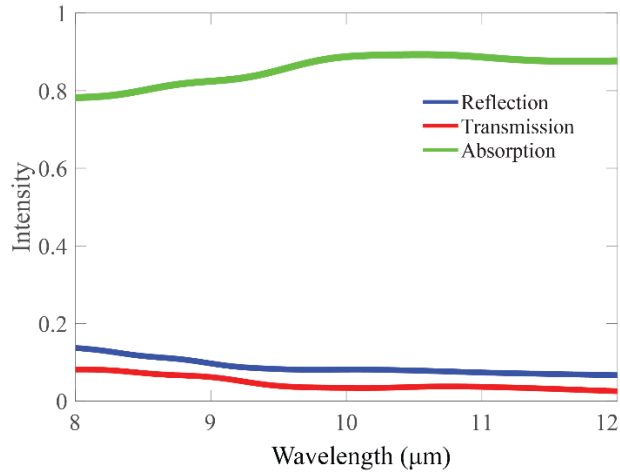

**Fig. 12** Reflection (R), transmission (T), and absorption (A) spectra of a metal-dielectric grating pixel structure, with 10 subwavelength grid unit cells (i.e., N=10) inside a linear period and a 2 μm gap ($g_t$) between adjacent pixels. For each grid unit cell, grid unit period $P_p$=1 μm, duty cycle $f_p$=0.2, $d_{Ti}$=50 nm, and $d_{SN}$= 170 nm are used.

Fig. 12 shows the simulated reflection (R), transmission (T), and absorption (A) of the grating pixel period with normally incident light and grid unit parameters specified as: $f_p$=0.2, $P_p$=1 μm, $d_{Ti}$=50 nm, and $d_{SN}$=170 nm. From our theoretical analysis, 10 grid unit cells (N=10) are considered inside the 1D pixel period (i.e., 100 cells inside the pixel area), with a periodic air gap $g_t$ of 2 μm. With these parameters, the pixel structure corresponds to a total area of 144 μm² with a filled area of 36 μm², i.e., a fill factor of 25%. From Fig. 12, it is evident that the grating pixel realizes excellent broadband absorption in between 8−12 μm, yielding an average absorption of ~86% with maximum absorption of ~90%.

## 4. Optimization of absorption in grating pixel

In general, guided mode excitation largely depends on the proper optimization of materials present in the grating structure [35, 60, 65]. In order to study optimum absorption in our pixel, we can consider the duty cycle $f_p$=0.2 (keeping minimum materials) and vary the grid unit period $P_p$. Fig. 13 shows the LWIR absorption profile with constant duty cycle $f_p$=0.2, and with varying grid unit periods ranging from 0.5 µm to 1.5 µm. The other parameters (i.e., N=10 and $g_t$=2 µm) are kept fixed as before. From Fig. 13, it can be observed that maximum absorption occurs when grid unit period $P_p$ is 1 µm. This can be attributed to the maximum coupling efficiency associated with the optimum pixel thickness. Our calculations indicate that a pixel thickness of 220 nm can realize guided mode coupling with air gap $g_t$=2 µm (Fig. 8), when grid unit parameters are: $f_p$=0.2 and $P_p$=1 µm. Therefore, $P_p$ larger or smaller than 1 µm cannot provide optimum absorption, which validates the theoretical analysis. To further illustrate this, Fig. 14 presents the spatial electric field distribution of pixel structure with grid unit periods of 0.5 µm, 1 µm and 1.5 µm. With the same parameters as Fig. 13, it can be observed that the evanescent field is coupled within the air gaps and maximum coupling occurs for period $P_p$=1 µm, realizing enhanced broadband absorption in the bilayer pixel by effectively extending the pixel size (absorption area) to encompass the open gaps between and around the pixels.

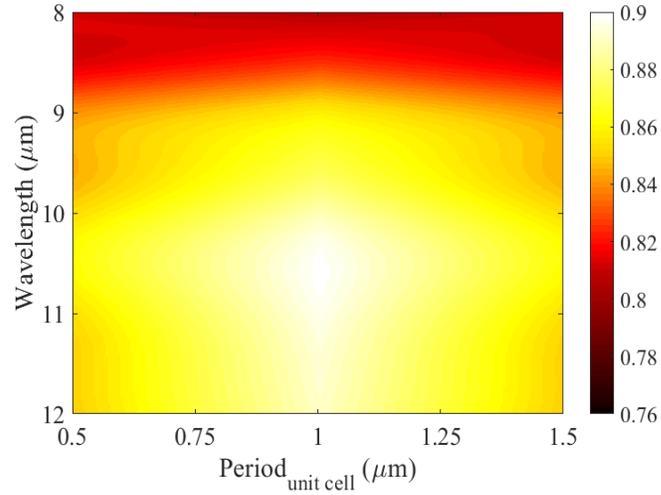

**Fig 13:** Absorption profile of a grating pixel structure with different periods of the grid unit cells. The duty cycle of the unit cells ($f_p$) is fixed at 0.2. The other specifications are: N=10 and $g_t$=2 μm.

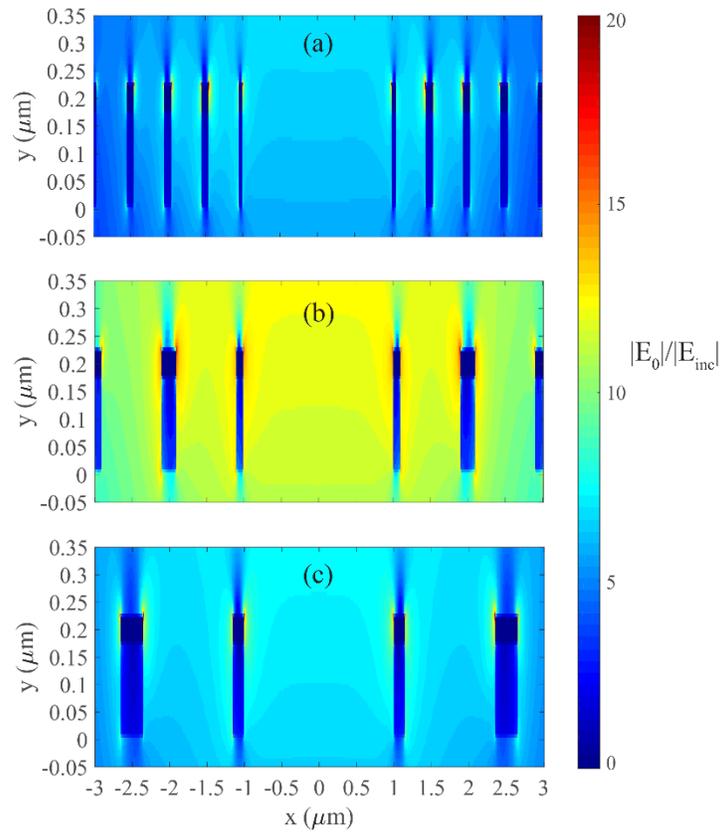

Fig 14: Spatial electric field distribution for the periodic grating pixel at λ=10.5 μm with different grid unit cell periods: (a) $P_p$=0.5 μm, (b) $P_p$=1 μm, and (c) $P_p$=1.5 μm. The duty cycle of the unit cells ($f_p$) is fixed at 0.2. The other specifications are: N=10 and $g_t$=2 μm.

## 5. Performance Analysis with absorption per unit mass

To understand optimum coupling with a minimum amount of material, we perform a comparative analysis of our pixel structure with the previously reported LWIR broadband absorbers. It should be noted that when an absorber period (e.g., pixel or periodic structure) is much larger than the operating wavelength (i.e., period>>λ), the total amount of material proportionately increases with the area, while the total absorption may not increase significantly [82, 83]. Therefore, to analyze the overall quality of the absorber, the absorption may need to be normalized to both mass and area. However, in case of an absorber period comparable to or smaller than the operating wavelength (i.e., period ≤ λ), the absorption per unit mass (absorption/mass) of the pixel or periodic structure would better define the absorber performance. Our proposed absorber, as well as the previously reported LWIR absorbers (analyzed in this section), have pixels or periodic structures smaller or comparable to the LWIR wavelengths (i.e., 8−12 µm) [4, 17, 24−26]. Therefore, we consider the absorption per unit mass as our performance metric.

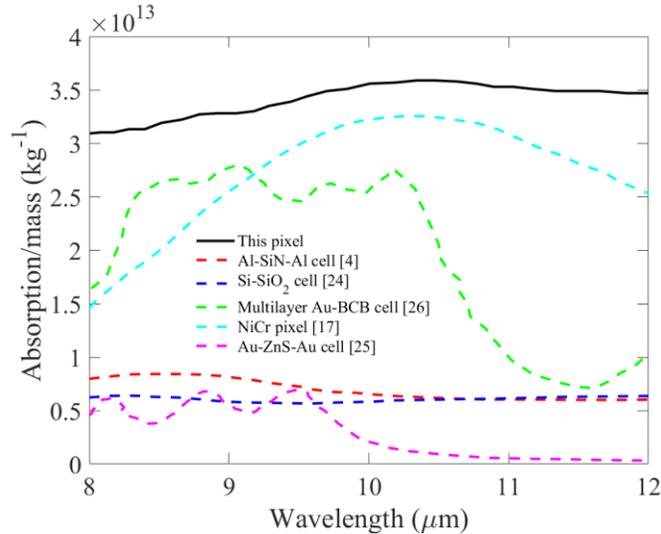

**Fig. 15** Comparative analysis of absorption/mass per period of our pixel structure with the state-of-the-art LWIR broadband absorbers, including metamaterials, metasurface and plasmonic structures.

**Table 1** Average absorption in the LWIR (8−12 μm) and total amount of mass of the absorber periods (pixels or periodic cells) from this and several previously reported works.

| Absorber | Average Absorption | Mass (kg) | Reference |
|---|---|---|---|
| This pixel | 86% | $2.5225\times10^{-14}$ | |
| Al-SiN-Al cell | 85% | $1.1858\times10^{-13}$ | [4] |
| Si-SiO$_2$ cell | 90% | $1.4862\times10^{-13}$ | [24] |
| Multilayer Au-BCB cell | 71% | $3.1178\times10^{-14}$ | [26] |
| NiCr pixel | 78% | $2.996\times10^{-14}$ | [17] |
| Au-ZnS-Au cell | 65% | $1.3818\times10^{-13}$ | [25] |

To compare the previously reported works, we consider all kind of structures, including metamaterials, metasurfaces and plasmonic structures. For example, Bouchon et al. [25] reported a wideband omnidirectional infrared absorber using an Au-ZnS-Au (M-I-M) structure, which includes four different sized resonators in a single period and realizes absorption through coupling multiple resonance bands. Later on, Adomanis et al. [26] proposed a fully functional bilayer metamaterial nearly perfect absorber, which consists of two pairs of Au-benzocyclobutene (BCB) films along with an Au substrate (i.e., Au-BCB-Au-BCB-Au, double M-I-M configuration). Gorgulu et al. [24] reported an ultra-broadband plasmonic infrared absorber, which considers symmetric 2D Si gratings on top of a SiO$_2$ layer. Jung et al. [17] presented the non-resonant broadband LWIR absorption in a metasurface pixel, formed by a NiCr film with periodic air holes. Moreover, Ustun et al. [4] reported broadband absorption in an Al-SiN-Al periodic structure (M-I-M) through a rigorous numerical analysis. To illustrate the amount of materials required for the broadband absorption, Table 1 shows the average absorption in the LWIR (i.e., 8−12 μm) and total amount of mass utilized in each of the absorber periods. Note that most LWIR broadband absorbers, specifically metamaterials and plasmonic composites, do not report pixel size or array geometry but rather focus on the details of the periodic cell structures. Therefore, in those cases we consider the reported periodic cells as the absorber periods for our analysis. From the table, it can be observed that the design methodology of this paper realizes an average absorption of ~86% with a mass amount of $2.5225\times10^{-14}$ kg for the pixel. This mass is at least ~18.77% smaller than the previously reported absorber periods. The use of only two layers (Ti and Si$_3$N$_4$) instead of tri- or multilayers and the use of optimum pixel design to increase evanescent field coupling within the gaps (due to GMR), within and around the pixel yield the enhanced broadband absorption with the minimum amount of materials. To further illustrate this, we calculate the absorption/mass per period (pixel or periodic cell) of the absorber structures in between 8−12 μm, which is shown in Fig. 15. From the figure, it can be observed that metamaterial with an M-I-M configuration, e.g., Al-SiN-Al, plasmonic structure e.g., Si-SiO$_2$, and plasmonic metamaterial e.g., Au-ZnS-Au show uniform absorption/mass (below $1\times10^{13}$ kg$^{-1}$) throughout the LWIR region. Due to having double M-I-M stacks, multilayer Au-BCB cell shows higher absorption/mass (~$2.5\times10^{13}$ kg$^{-1}$) in a smaller window (i.e., 8.3−10.2 μm). Comparatively higher but significantly nonuniform absorption/mass profile is observed for the NiCr pixel, which has a maximum value of $3.2\times10^{13}$ kg$^{-1}$. Our pixel structure shows a high and uniform absorption/mass profile with a maximum value of ~$3.6\times10^{13}$ kg$^{-1}$ near at ~10.5 μm (resonance) and with a standard deviation of only 4.2% across the entire LWIR. An average absorption/mass of $3.45\times10^{13}$ kg$^{-1}$ is found for our pixel, which is ~1.33−7.33 times larger than the previously reported LWIR absorbers.

## 6. Conclusion

In summary, we report that nearly perfect absorption across the LWIR can be possible with a minimum amount of material using pixel designs that incorporate guided-mode absorption resonances and optimize the evanescent field coupling in the air holes within and around the pixels. In effect, the overall absorption area per pixel becomes much larger than the amount of solid area used to create it. Lossy Ti-Si$_3$N$_4$ waveguide gratings with low fill factors and periodic open gaps are used to form the broadband pixel. Through a theoretical analysis, we optimize the thickness of the pixel structure (along with grating or grid units) to realize guided mode resonance conditions. Using finite difference time domain technique (FDTD), we calculate the broadband LWIR absorption of the pixel structure. By using lossy metal-dielectric bi-layers and utilizing evanescent field coupling inside the holes and gaps (due to GMR), excellent broadband absorption is observed for the optimized pixel, with an average absorption of ~86% and maximum absorption of ~90% in between the wavelengths of 8−12 μm. The total mass required for our pixel is found $2.5225\times10^{-14}$ kg, which is at least ~18.77% smaller than the previously reported absorber periods. The reduction of mass results in an enhanced absorption/mass per period with an average of ~$3.45\times10^{13}$ kg$^{-1}$, which is ~1.33−7.33 times larger than the state-of-the-art LWIR absorbers.

**Funding**
The authors gratefully acknowledge funding from the Army Research Office under grant W911-NF-18-1-0272.

**Disclosures**
The authors declare no conflicts of interest.

**See Supplement 1 for supporting content**